\documentclass[10pt,conference]{IEEEtran}
\usepackage{epsfig,setspace,amsmath,epsf,amssymb,bm,theorem,cite,graphicx, epstopdf,algorithm,float,color,mathtools, authblk}
\usepackage[table,xcdraw]{xcolor}
\usepackage{subcaption}
\usepackage{algpseudocode}
\usepackage{dsfont}
\usepackage[short,c2]{optidef}
\usepackage{amsmath}
\usepackage{physics}

\DeclareMathOperator*{\argmax}{arg\,max}

\newtheorem{theorem}{Theorem}
\newtheorem{corollary}{Corollary}

\newtheorem{remark}{Remark}
\newtheorem{lemma}{Lemma}

\newcommand{\e}{{\mathbb{E}}}

\IEEEoverridecommandlockouts
\allowdisplaybreaks

\begin{document}

\title{Age of Estimates: When to Submit Jobs to a Markov Machine to Maximize Revenue}

\author{Sahan Liyanaarachchi \qquad Sennur Ulukus\\
	\normalsize Department of Electrical and Computer Engineering\\
	\normalsize University of Maryland, College Park, MD 20742 \\
	\normalsize \emph{sahanl@umd.edu} \qquad \emph{ulukus@umd.edu}}

% \author{\hspace*{1cm}\\
% 	\normalsize \hspace*{1cm}\\
% 	\normalsize \hspace*{1cm} \\
% 	\normalsize \hspace*{1cm}}    

\maketitle

\begin{abstract}
    With the dawn of AI factories ushering a new era of computing supremacy, development of strategies to effectively track and utilize the available computing resources is garnering utmost importance. These computing resources are often modeled as Markov sources, which oscillate between free and busy states, depending on their internal load and external utilization, and are commonly referred to as Markov machines (MMs). Most of the prior work solely focuses on the problem of tracking these MMs, while often assuming a rudimentary decision process that governs their utilization. Our key observation is that the ultimate goal of tracking a MM is to properly utilize it. In this work, we consider the problem of maximizing the utility of a MM, where the utility is defined as the average revenue generated by the MM. Assuming a Poisson job arrival process and a query-based sampling procedure to sample the state of the MM, we find the optimal times to submit the available jobs to the MM so as to maximize the average revenue generated per unit job. We show that, depending on the parameters of the MM, the optimal policy is in the form of either a \emph{threshold policy} or a \emph{switching policy} based on the \emph{age of our estimate} of the state of the MM.
\end{abstract}

\section{Introduction}
In the midst of the AI revolution, a significant amount of resources have been directed towards the development of high-end compute units in the modern world. As the power and the density of these commercially available compute units increase, what ultimately will decide one's competitive edge over the others, is the ability to effectively utilize the available computing power for the required task. Hence, effectively tracking and utilizing the compute units have gained significant attention in the recent literature. 

Most of the existing literature mainly focuses on only one of the tasks listed above: tracking the state of the compute units. In particular, most works are concentrated around the development of sampling strategies to closely track the status of the compute units \cite{graves2024, subhankar_MM, Markov_machines, melih_MM}. In these works, the compute units are often modeled as Markov sources where freshness based metrics such as the binary freshness metric (BFM) \cite{melih_BF_cache, melih_BF_gossip, melih_IF_CUS, melih_BF_Inf, structured_estimators}, age of information (AoI) \cite{yates2020age, age1, age2} and age of incorrect information (AoII) \cite{AoII2019, AoII_Markov, ismail_AoII, ismail_map} have been used to evaluate the effectiveness of the tracking procedure. However, the ultimate goal of tracking these Markovian compute units, also known as Markov machines (MMs), is to effectively utilize them. Therefore, development of control policies that directly maximize the utility of the MMs, rather than freshness of their state estimates, is of significant interest.

\begin{figure}[t]
    \centering
    \includegraphics[width=\columnwidth]{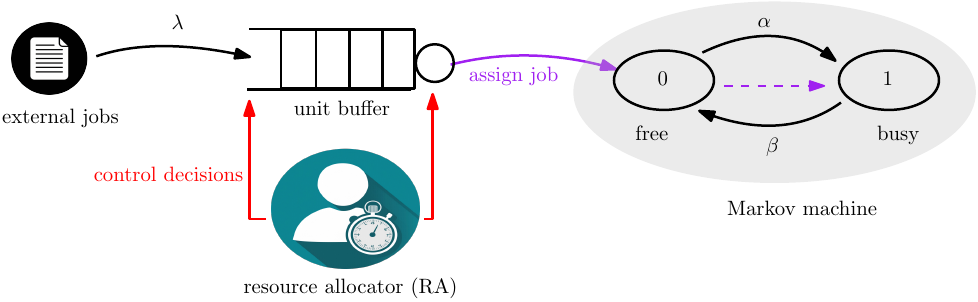}
    \caption{System with waiting-based job assignment.}
    \label{fig:sys_model}
\end{figure}

In this work, we explicitly study the utility maximization problem for these MMs. In particular, we consider a MM which oscillates between free and busy states according a binary continuous-time Markov chain (CTMC). This MM is subjected to a query-based sampling procedure, where a resource allocator (RA) sends queries to inquire about the state of the MM at exponentially distributed time intervals. At the same time, a Poisson arrival process brings in job requests to the RA, where the RA, based on its current estimate of the MM, decides when to submit these job requests to the MM; see Fig.~\ref{fig:sys_model}. When a job is submitted to the MM, the MM may be busy or free in actuality. In the former case, a penalty will be paid, whereas a revenue will be generated in the latter case. Hence, the job submission times must be carefully chosen. 

In this work, we define the utility of the MM as the revenue generated by the MM and find the optimal job submission times to maximize its utility. In particular, we decide on how long the RA should wait before submitting the job depending on the \emph{value} of its current estimate of the MM (i.e., free or busy) and the \emph{age of its current estimate}. We show that these optimal wait times form either a \emph{threshold policy} (wait to submit) or a \emph{switching policy} (submit or never submit) based on the age of the estimate at the RA.

\section{Related Work}
The query-based sampling procedure was introduced in \cite{nail_QS}, where the authors look into the optimal rate allocation problem for monitoring multiple heterogeneous CTMCs. In \cite{nail_QS}, the authors maximize the binary freshness metric (BFM) which penalizes when there is a disparity between the estimate and the actual state of the CTMC. The work in \cite{graves2024} looks into the problem of allocating and monitoring workers in a distributed computation setting, where each worker is modeled as a binary discrete time source with free and busy states. In \cite{graves2024}, each source is assigned a different priority level and the authors find the optimal update pattern that should be followed by these sources so as to minimize the error probability of the top-$k$ set of these sources.

The work in \cite{melih_MM} considers the problem of allocating tasks to workers so as to maximize their efficiency. In \cite{melih_MM}, the workers are modeled as CTMCs but with multiple states, each corresponding to a different level of efficiency for the worker. The authors show that, under certain conditions, the optimal sampling rate of these workers is in the form a threshold policy. The work in \cite{subhankar_MM} looks into the problem of tracking and assigning jobs into a MM in the discrete time setting. In \cite{subhankar_MM}, the MM shifts between free and busy states while a sampler at the MM decides when to pull jobs from a server. They too consider a penalty for assigning a job when the machine is busy and optimize for a custom AoII metric.

The closest to our work is \cite{Markov_machines}, where the authors consider the optimal rate allocation problem for tracking multiple MMs with job assignment. They show that age based metrics do not necessarily improve the utility of the MMs, and they introduce two new metrics to evaluate how the sampling rate affects the efficacy of the underlying decision process. However, their considered decision process is rather naive. In \cite{Markov_machines}, a job is immediately submitted to the MM or discarded by the RA upon a job arrival. In our work, we directly optimize for the utility of the MMs and show that waiting before submitting jobs is optimal for these MMs.

\section{System Model}
Internal jobs arrive at the Markov machine at a rate of $\alpha$ and any submitted job is processed by the MM at a service rate of $\beta$. Hence, the dynamics of the unadulterated MM can be modeled as a binary CTMC which oscillates between the free and busy states denoted by state 0 and state 1, respectively. Let $X(t)\in \{0,1\}$ denote the actual state of the MM.  This MM is sampled via queries which report back the state of the MM instantaneously to the RA. The RA will retain this state as its estimate until it is updated again. Let $\mu$ be the rate of sampling and let $\hat{X}(t)\in\{0,1\}$ represent the estimate of the MM held by the RA. 

\begin{figure}[t]
    \centering
    \includegraphics[width=0.5\columnwidth]{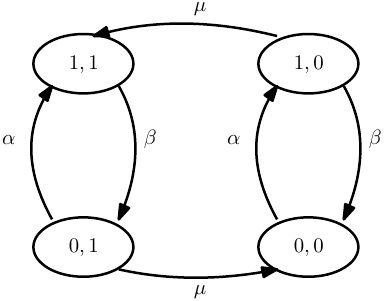}
    \caption{Query-based sampling process with state space $(X(t),\hat{X}(t))$.}
    \label{fig:query_MC}
\end{figure}

Now, the dynamics of the query based sampling procedure can be represented by a two-dimensional Markov chain whose states are $(X(t),\hat{X}(t))$. Due to the internal dynamics of the MM, until we sample, this CTMC will oscillate between either states $(0,0)$ and $(1,0)$, or states $(0,1)$ and $(1,1)$, each with rates $\alpha$ and $\beta$, respectively. If we sample in state $(0,1)$, the chain will transition to state $(0,0)$ and if will sample in state $(1,0)$, the chain will transition to state $(1,1)$. Each of these transitions will take place with rate $\mu$. The dynamics of the query-based sampling procedure is illustrated in Fig.~\ref{fig:query_MC}.

External job requests arrives at the RA at a rate of $\lambda$. We assume that, once an external job request arrives at the RA, the RA will hold the job in a unit-sized buffer until it is submitted to the MM. Any external job requests that arrive at the RA while the RA is holding onto a job (buffer occupied) will be lost to the RA. When the RA submits a job, if the MM is actually free, then this external job request will be processed by the MM generating a revenue of $r_s$. However, at the time of job submission, if the MM is actually busy, then the submitted job will be discarded by the MM incurring a penalty of $c_d$. In either case, the action of submitting a job will unveil the actual state of the MM to the RA. In particular, at the time of job submission, $\hat{X}(t)$ will always be updated to $\hat{X}(t)=1$. Therefore, right after job submission, the query based sampling procedure will always start from state $(1,1)$.

When an external job arrives at the RA, its estimate $\hat{X}(t)$ may be outdated from the last time it received a sample or submitted a job. Let $\Delta$ denote the time elapsed since that last time the RA was aware of the exact status of the MM. Now, when there is a job present at the RA, the task of the RA is to decide when to submit the job based on $\Delta$ and $\hat{X}(t)$ so as to maximize the average revenue generated per job given by,
\begin{align}
    R=\e\left[\lim_{t\to\infty}\frac{r_s S(t)-c_dD(t)}{N(t)}\right],
\end{align}
where $S(t)$ is the number of jobs submitted successfully to the MM, $D(t)$ is the number of jobs discarded at the MM with a penalty and $N(t)$ is the total number of jobs that arrived at the RA by time $t$. We further consider only stationary policies (decisions depends only on $\hat{X}(t)$ and $\Delta$) and hence this limit exists almost surely. Therefore, the goal of this work is to find the optimal stationary policy that maximizes $R$.
\begin{remark}
    The time averaged revenue is equal to $\lambda R$, hence, any policy that maximizes $R$, also maximizes the average revenue generated per unit time.
\end{remark}

\section{Main Results}
When a job arrives at the RA when the buffer is empty, that job is accepted and the RA decides when to submit the job to the MM. At this time instance, suppose $\hat{X}(t)=i$ and $\Delta=u$. Then, based on $u$ and $i$, the RA decides when to submit this job. Let $\tau_{i,u}$ denote this time. Then, let $R(i,u,\tau_{i,u})$ be the expected revenue generated and $L(i,u,\tau_{i,u})$ be the expected number of jobs lost at the RA when employing the decision $\tau_{i,u}$. Then, Theorem \ref{thrm:avg_rev} below gives the expression for the average revenue in terms of the waiting times.

\begin{theorem}\label{thrm:avg_rev} 
    Under a stationary policy, the average revenue is given by,
    \begin{align}
        R=\frac{\sum_{i=0}^1p_i\e[R(i,U,\tau_{i,U})]}{1+\sum_{i=0}^1p_i\e[L(i,U,\tau_{i,U})]},
    \end{align}
    where $p_i$ is the probability of accepting a job when $\hat{X}(t)=i$ and $U\sim Exp(\lambda+\mu)$.
\end{theorem}

The proof of Theorem \ref{thrm:avg_rev} is given in Appendix \ref{apen:thrm_avg_rev}. 

Next, we need to find the set of waiting times $\tau_{i,u}$ when $\hat{X}(t)=i$ and $\Delta=u$ that maximizes $R$. This results in the following optimization problem,
\begin{align}
\theta^*=\max_{\tau_{i,u}\geq0}\frac{\sum_{i=0}^1p_i\e[R(i,U,\tau_{i,U})]}{1+\sum_{i=0}^1p_i\e[L(i,U,\tau_{i,U})]}.
\end{align}

To solve the above problem, we employ Dinkelbach's method to linearize the fractional problem\cite{dinkelbach}. To proceed with our analysis, define the linearized optimization problem as follows,
\begin{align}
    J(\theta)=& \max_{\tau_{i,u}\geq0} \Bigg\{\sum_{i=0}^1p_i\e[R(i,U,\tau_{i,U})]\nonumber\\
    &\qquad\qquad-\theta(1+\sum_{i=0}^1p_i\e[L(i,U,\tau_{i,U})])\Bigg\}\nonumber\\
    =& \max_{\tau_{i,u}\geq0}\sum_{i=0}^1p_i\left(\e[R(i,U,\tau_{i,U})]-\theta\e[L(i,U,\tau_{i,U})]\right)-\theta\nonumber\\
    =&\max_{\tau_{i,u}\geq0}\sum_{i=0}^1p_i\e[V_\theta(i,U,\tau_{i,U})]-\theta,
\end{align}
where $V_{\theta}(i,u,\tau_{i,u})=R(i,u,\tau_{i,u})-\theta L(i,u,\tau_{i,u})$.

Then, the following relation holds true: $J(\theta)\lesseqgtr0\iff\theta^*\lesseqgtr\theta$. Hence, to find $\theta^*$, we can simply find the optimal waiting times for $J(\theta)$ and find the optimal $\theta$ using a bisection search. Moreover, since the external job arrival rate is $\lambda$, we have $L(i,u,\tau_{i,u})=\lambda\tau_{i,u}$. Therefore, $\theta$ can be viewed as the indirect holding cost of job request at the RA.

\begin{remark}
    Note that, when $\theta=0$, the naive policy where we submit jobs only when $i=\Delta=0$ yields that $J(0)>r_s>0$. Hence, $\theta^*>0$. Therefore, we only consider $\theta>0$.
\end{remark}

Let $P(t)$ denote the transition probabilities of the unadulterated CTMC at time $t$. Then, $P(t)$ will be given by,
\begin{align}
    P(t)=&\begin{bmatrix}
    P_{00}^t&P_{01}^t\\
     P_{10}^t&P_{11}^t
    \end{bmatrix}\\
    =& \begin{bmatrix}
    \frac{\beta}{\alpha+\beta}+\frac{\alpha}{\alpha+\beta}e^{-(\alpha+\beta)t}&\frac{\alpha}{\alpha+\beta}-\frac{\alpha}{\alpha+\beta}e^{-(\alpha+\beta)t}\\
     \frac{\beta}{\alpha+\beta}-\frac{\beta}{\alpha+\beta}e^{-(\alpha+\beta)t}&\frac{\alpha}{\alpha+\beta}+\frac{\beta}{\alpha+\beta}e^{-(\alpha+\beta)t}
    \end{bmatrix}.\label{eqn:P_mat}
\end{align}
Now, using $P(t)$, we can find the optimal wait times as stated by Theorem \ref{thrm:opt_wait}.

\begin{theorem}\label{thrm:opt_wait}
    The optimal wait times, $\tau^*_{i,u}$ that minimizes $J(\theta)$ are given by,
    \begin{align}
        \tau^*_{i,u}
        =&\argmax_{\tau\geq0}\bigg\{(r_sP_{i0}^{u+\tau}-c_dP_{i1}^{u+\tau})e^{-\mu\tau}-\theta\lambda\frac{(1-e^{-\mu\tau)}}{\mu}\nonumber\\
        &\qquad\qquad \ +\sum_{j=0}^1\e\left[P_{ij}^{u+Y}\mathds{1}\{Y\leq\tau\}
        \right]V_j\bigg\},\label{eqn:opt_times_0}
    \end{align}
    for $u>0$, where $Y\sim Exp(\mu)$, $V_0=r_s$ and $V_1$ is given by,
    \begin{align}
        V_1=&\max_{\tau\geq 0}\Bigg\{\frac{(r_sP_{10}^{\tau}-c_dP_{11}^{\tau})e^{-\mu\tau}-\theta\lambda\frac{(1-e^{-\mu\tau)}}{\mu}}{1-\e\left[P_{11}^{Y}\mathds{1}\{Y\leq\tau\}
        \right]}\nonumber\\
        &\qquad\quad+\frac{r_s\e\left[P_{10}^{Y}\mathds{1}\{Y\leq\tau\}
        \right]}{1-\e\left[P_{11}^{Y}\mathds{1}\{Y\leq\tau\}
        \right]}\Bigg\}.\label{eqn:tau_1}
    \end{align}
    Moreover, $\tau^*_{0,0}=0$ and $\tau^*_{1,0}$ is the maximizer of \eqref{eqn:tau_1}.
\end{theorem}
\begin{corollary}\label{cor:opt_times_1}
    The optimal waiting times $\tau^*_{i,u}$ for $u>0$, that minimize $J(\theta)$, are given by,
    \begin{align}
        \tau^*_{0,u}=&\argmax_{\tau\geq0}\Bigg\{Ae^{-\mu\tau}+e^{-(\alpha+\beta)u}B_0e^{-(\alpha+\beta+\mu)\tau}\Bigg\},\\
        \tau^*_{1,u}=&\argmax_{\tau\geq0}\Bigg\{Ae^{-\mu\tau}-e^{-(\alpha+\beta)u}B_1e^{-(\alpha+\beta+\mu)\tau}\Bigg\},
    \end{align}
    where  $B_0=\frac{\alpha}{\alpha+\beta}B$ and $B_1=\frac{\beta}{\alpha+\beta}B$ with $A$ and $B$ defined as follows,
    \begin{align}
        A&=\frac{\lambda\theta}{\mu}-\frac{(c_d+V_1)\alpha}{\alpha+\beta},\\
        B&=c_d+\frac{(\alpha+\beta)r_s+\mu V_1}{\alpha+\beta+\mu}.
    \end{align}
\end{corollary}

The proof of Theorem \ref{thrm:opt_wait} is presented in Appendix \ref{apen:thrm_opt_wait} and the proof of Corollary \ref{cor:opt_times_1} follows directly by substituting \eqref{eqn:P_mat} in \eqref{eqn:opt_times_0} and discarding the constant terms. 

Now, the next couple of theorems completely characterize the optimal waiting times for $J(\theta)$.

\begin{theorem}\label{cor:opt_times_2}
    If $A>0$, then the optimal wait times are given by,
    \begin{align}
        \tau^*_{0,u}&=0,\\
        \tau^*_{1,u}&= \Big(\frac{1}{\alpha+\beta}\ln{\frac{(\alpha+\beta+\mu)B_1}{\mu A}}-u\Big)^+,
    \end{align}
    where $(x)^+=\max\{0,x\}$.
\end{theorem}

\begin{theorem}\label{cor:opt_times_3}
    If $A\leq0$, then the optimal wait times are given by,
    \begin{align}
        \tau^*_{0,u}&=\begin{cases}
            0&\text{if}~\quad u\leq \frac{1}{\alpha+\beta}(\ln{\frac{B_0}{|A|}})^+\\
            \infty&\text{if}~\quad u> \frac{1}{\alpha+\beta}(\ln{\frac{B_0}{|A|}})^+,\\
        \end{cases}\\
        \tau^*_{1,u}&=\infty.
    \end{align}
\end{theorem}

The proofs of Theorems \ref{cor:opt_times_2} and \ref{cor:opt_times_3} are given in Appendix \ref{apen:cor_opt_times}. In here, $\tau^*_{i,u}=\infty$ implies that, we simply wait for a new update on the state of the MM to come to the RA for the RA to decide when to submit the job. Note that, Theorem \ref{cor:opt_times_2} states that the optimal wait times follow a threshold policy, and Theorem \ref{cor:opt_times_3} states that the optimal wait times follow a switching policy, both based on the age of the estimate $u$.

Next, we find $J(\theta)$ using these optimal waiting times.

\begin{lemma}\label{lem:avg_rev_pos}
    For $A>0$,  $\e[V_\theta(i,U,\tau^*_{i,U})]$ will be given by,
    \begin{align}
        \e[V_\theta(0,U,\tau^*_{0,U})]&=\frac{\beta r_s-\alpha c_d}{\alpha+\beta}+\frac{\alpha(\mu+\lambda)(r_s+c_d)}{(\alpha+\beta)(\alpha+\beta+\mu+\lambda)},\\
        \e[V_\theta(1,U,\tau^*_{1,U})]&=\frac{\beta r_s-\alpha c_d}{\alpha+\beta}-a_0+a_1e^{-\mu\Gamma}-a_2e^{-(\lambda+\mu)\Gamma},
    \end{align}
where $\Gamma=\Big(\frac{1}{\alpha+\beta}\ln{\frac{(\alpha+\beta+\mu)B_1}{\mu A}}\Big)^+$, and $a_0$, $a_1$ and $a_2$ are defined as follows,
\begin{align}
    a_0&=\frac{\beta \mu(\lambda+\mu)(r_s-V_1)}{(\alpha+\beta)(\alpha+\beta+\mu)(\alpha+\beta+\mu+\lambda)}+A,\\
    a_1&=\frac{(\alpha+\beta)(\lambda+\mu)A}{\lambda(\alpha+\beta+\mu)},\\
    a_2&=\frac{\mu(\alpha+\beta)A}{\lambda(\alpha+\beta+\mu+\lambda)}.
\end{align}
\end{lemma}

\begin{lemma}\label{lem:avg_rev_neg}
    For $A\leq 0$,  $\e[V_\theta(i,U,\tau^*_{i,U})]$ is given by,
    \begin{align}
        \e[V_\theta(0,U,\tau^*_{0,U})]=&\frac{\beta r_s-\alpha c_d}{\alpha+\beta}+b_0-\frac{A(\alpha+\beta)e^{-(\lambda+\mu)\kappa}}{\alpha+\beta+\mu+\lambda},\\
        \e[V_\theta(1,U,\tau^*_{1,U})]=&\frac{\beta r_s-\alpha c_d}{\alpha+\beta}-A\nonumber\\
        &-\frac{\beta \mu(\lambda+\mu)(r_s-V_1)}{(\alpha+\beta)(\alpha+\beta+\mu)(\alpha+\beta+\mu+\lambda)},
    \end{align}
    where $\kappa=\frac{1}{\alpha+\beta}(\ln{\frac{B_0}{|A|}})^+$ and $b_0$ is defined as,
    \begin{align}
        b_0=\frac{\alpha(\lambda+\mu)(r_s+c_d)}{(\alpha+\beta)(\alpha+\beta+\mu+\lambda)}.
    \end{align}
\end{lemma}

The proofs of Lemma \ref{lem:avg_rev_pos} and Lemma \ref{lem:avg_rev_neg} can be found in Appendix \ref{apen:lem_avg_rev_pos} and Appendix \ref{apen:lem_avg_rev_neg}, respectively. 

\begin{figure}
    \centering
    \includegraphics[width=\columnwidth]{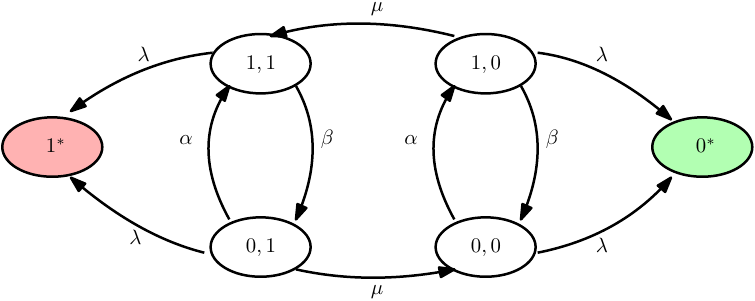}
    \caption{Absorbing Markov chain $Z(t)$.}
    \label{fig:amc}
\end{figure}

Now, all that is left is to find the $p_i$s to fully characterize $J(\theta)$. To find the $p_i$s, we first construct the absorbing Markov chain $Z(t)$ illustrated in Fig.~\ref{fig:amc}. The $p_i$s will be simply given by,
\begin{align}
    p_i=\mathds{P}(Z(\infty)=i^*|Z(0)=(1,1)).
\end{align}

Let the states of $Z(t)$ be ordered as $(1,1)$, $(0,1)$, $(0,0)$, $(1,0)$, $0^*$, $1^*$ and let $Q$ be its generator matrix with respect to the same ordering. Then, $Q$ will be in the following form,
\begin{align}
    Q=\begin{bmatrix}
        \Lambda&\Psi\\
        0_{2\times 4}&0_{2\times 2}
    \end{bmatrix},
\end{align}
where $\Lambda$ is the $4\times4$ sub generator matrix corresponding to the transient states of $Z(t)$, $\Psi$ is the $4\times 2$ matrix whose elements are the transition rates from transient states to absorbing states, $0_{2\times 4}$ is a $2\times4$ matrix of zeros and $0_{2\times 2}$ is a $2\times2$ matrix of zeros. Then, the next lemma gives us the $p_i$ values in terms of $\Lambda$ and $\Psi$.

\begin{lemma}\label{lem:pi}
    The $p_i$ values are given by the following expression,
    \begin{align}
        \begin{bmatrix}
            p_0\\
            p_1
        \end{bmatrix}=-\bm{v}_0^T\Lambda^{-1}\Psi,
    \end{align}
    where $\bm{v}_0$ is a vector of all ones except at the first location. 
\end{lemma}

The proof of Lemma \ref{lem:pi} is presented in Appendix \ref{apen:lem_pi}. 

With this, we can find $J(\theta)$ and then the optimal value for $\theta$ can be found using a simple bisection search. In particular, if $J(\theta)>0$, then we increase $\theta$, and decrease it otherwise, until we reach a stable $\theta$ value within a desired error bound. 

\begin{remark}\label{rem:pi}
    The $p_i$s are independent of the decision process since each time we submit a job, $\hat{X}(t)=X(t)=1$, and the decision process will become active only when there is a job present at the RA for submission to the MM.
\end{remark}

\begin{figure*}
    \centering
    \begin{subfigure}[b]{0.45\textwidth} % [position]{width}
        \centering
        \includegraphics[width=\textwidth]{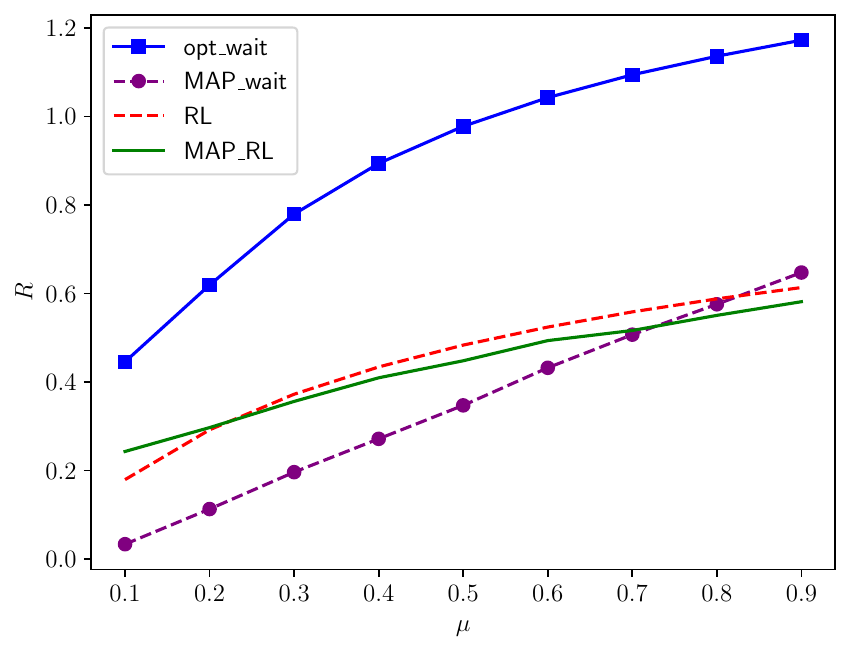}
        \caption{$\alpha=0.2$, $\beta=0.5$.}
        \label{fig:var_mu_low_alpha}
    \end{subfigure}
    \hfill 
    \begin{subfigure}[b]{0.45\textwidth}
        \centering
        \includegraphics[width=\textwidth]{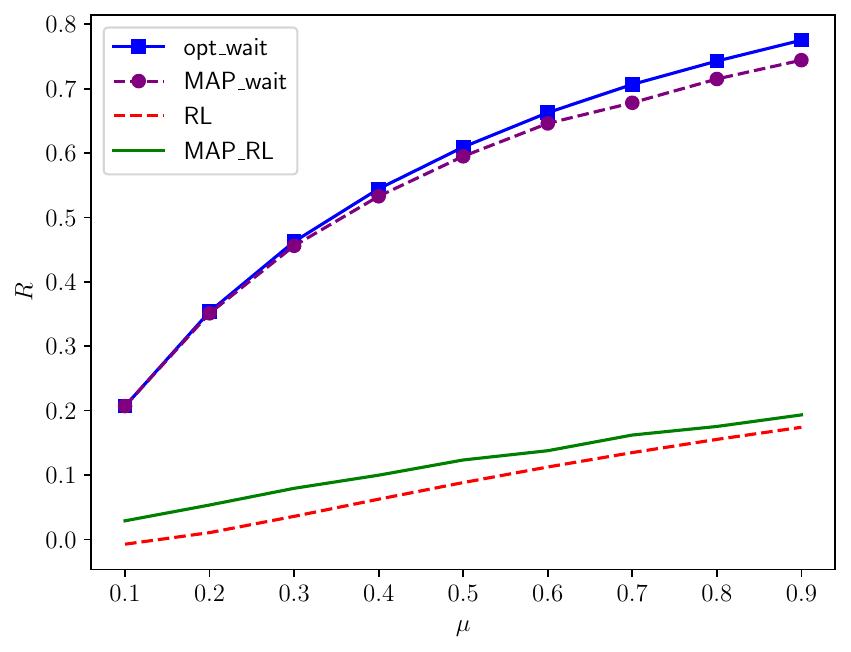}
        \caption{$\alpha=0.5$, $\beta=0.3$.}
        \label{fig:var_mu_high_alpha}
    \end{subfigure}
    \vspace*{0.2cm}
    \caption{Variation of average revenue per unit job $R$ with $\mu$ for $\lambda=0.3$, $r_s=2$ and $c_d=3$ across two MMs.}
    \label{fig:var_mu}
\end{figure*}

\begin{figure*}
    \centering
    \begin{subfigure}[b]{0.45\textwidth} % [position]{width}
        \centering
        \includegraphics[width=\textwidth]{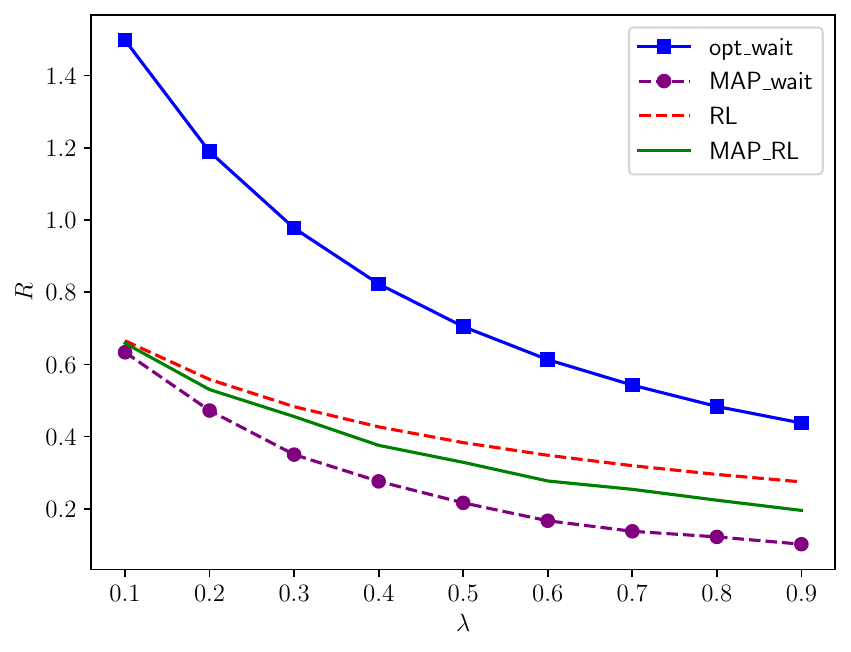}
        \caption{$\alpha=0.2$, $\beta=0.5$.}
        \label{fig:var_lambda_low_alpha}
    \end{subfigure}
    \hfill 
    \begin{subfigure}[b]{0.45\textwidth}
        \centering
        \includegraphics[width=\textwidth]{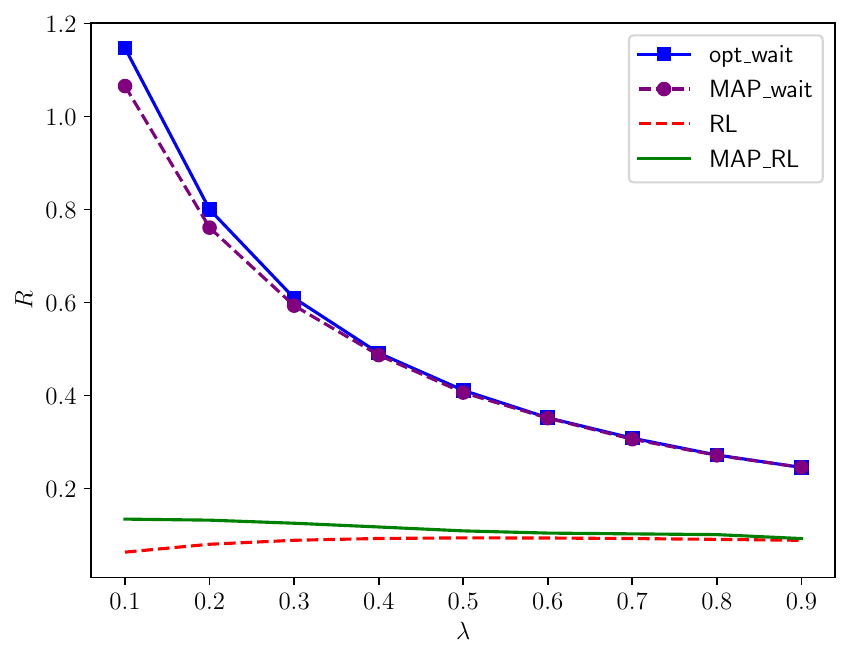}
        \caption{$\alpha=0.5$, $\beta=0.3$.}
        \label{fig:var_lambda_high_alpha}
    \end{subfigure}
    \vspace*{0.2cm}
    \caption{Variation of average revenue per unit job $R$ with $\lambda$ for $\mu=0.5$, $r_s=2$ and $c_d=3$ across two MMs.}
    \label{fig:var_lambda}
\end{figure*}

\section{Numerical Results}
In this section, we compare our optimal policy which we term \emph{opt\_wait}, with several benchmark policies. The considered benchmark policies are the following:
\begin{enumerate}\itemsep1em
    \item \emph{RL}: This is the naive \emph{restless} job submission policy considered in \cite{Markov_machines} where the RA immediately submits jobs or discards jobs upon a job arrival based on its current estimate of the state of the MM. In particular, if $\hat{X}(t)=0$ when a job arrives at the RA, it will be immediately submitted to the MM by the RA. If $\hat{X}(t)=1$ on the other hand, the job will not be accepted and will be discarded by the RA.
    
    \item \emph{MAP\_RL}: This is an extension of the RL policy where the RA submits the job to the MM immediately upon its arrival based on the MAP estimate of the state of the MM at the RA. In particular, when a job arrives at the RA, based on the age of the estimate, the RA will compute the most likely state of the MM at that time instance. If the MM is more likely to be free, then the job is submitted. Otherwise, the job is not accepted and is discarded by the RA.
    
    \item \emph{MAP\_wait}: In this policy, similar to our \emph{opt\_wait} policy, the RA will hold a job for sometime before submitting to the MM. However, in here, the wait times are computed based on the MAP estimate of the MM. In particular, we will postpone  the job submission until the MAP estimate of the MM says that the MM is more likely to be free. At the same time, any incoming jobs while holding on to a job, will be lost to the RA. Let $\tau^M_{i,u}$ be the waiting times for this policy when $\hat{X}(t)=i$ and $\Delta=u$. Then, $\tau^M_{i,u}$ will be given by,
    \begin{align}
        \tau^M_{0,u}&=\begin{cases}
            0, & \text{if}~ \beta\geq\alpha ~\text{or}~ u\leq \frac{1}{\alpha+\beta}\ln{|\frac{2\alpha}{\alpha-\beta}|},\\
            \infty, & \text{otherwise},
        \end{cases}\\
        \tau^M_{1,u}&=\begin{cases}
            \infty, &\text{if}~\alpha\geq\beta,\\
            \left(\frac{1}{\alpha+\beta}\ln{|\frac{2\beta}{\beta-\alpha}|}-u\right)^+, & \text{otherwise}.
        \end{cases}
    \end{align}
\end{enumerate}

In the first experiment, we evaluate the variation of $R$ with $\mu$ for two different MMs: For the first MM, we consider that the internal job arrival rate is smaller compared to its processing rate, i.e., $\alpha<\beta$, and in the second MM we consider the opposite scenario, i.e., $\beta<\alpha$. As shown in Fig.~\ref{fig:var_mu}, in either case, \emph{opt\_wait} policy outperforms all the benchmarks. When $\alpha<\beta$, the performance gap between \emph{opt\_wait} and the rest is significantly higher (see Fig.~\ref{fig:var_mu_low_alpha}) whereas when $\beta<\alpha$, we see that MAP\_wait policy closely follows our optimum policy for low values of $\mu$ (see Fig.~\ref{fig:var_mu_high_alpha}). However, the performance gap between the two, tends to increase as $\mu$ is increased. 

In the second experiment, we evaluate how $R$ varies with $\lambda$. As shown in Fig.~\ref{fig:var_lambda}, our policy outperforms the benchmark policies in all cases. As before, we see that, when $\alpha>\beta$, the MAP\_wait policy is close to our optimal policy. However, the performance gap increases as $\lambda$ decreases (see Fig.~\ref{fig:var_lambda_high_alpha}). 

\begin{figure*}[t]
    \centering
    \begin{subfigure}[b]{0.45\textwidth} % [position]{width}
        \centering
        \includegraphics[width=\textwidth]{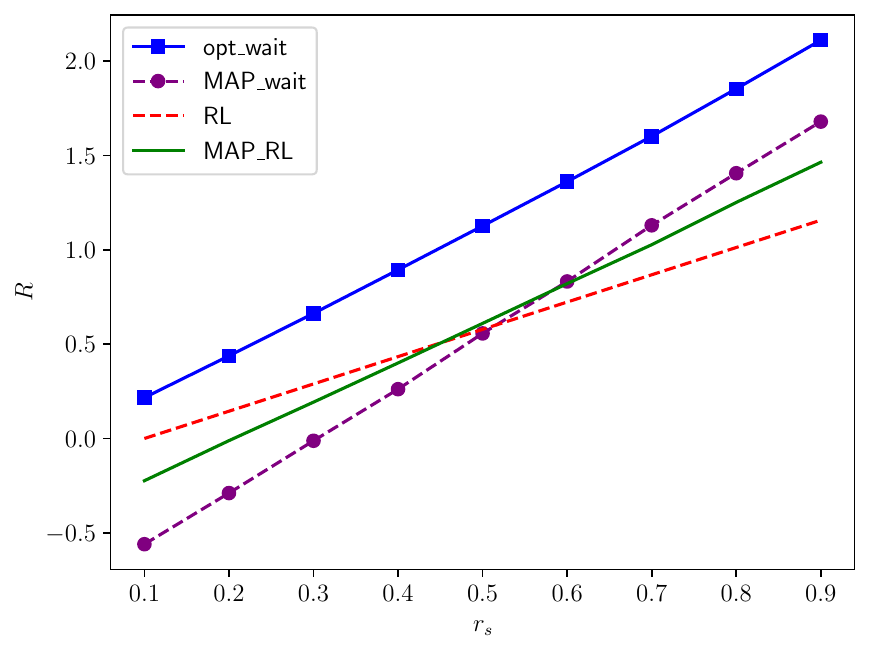}
        \caption{$\alpha=0.2$, $\beta=0.5$.}
        \label{fig:var_rs_low_alpha}
    \end{subfigure}
    \hfill
    \begin{subfigure}[b]{0.45\textwidth}
        \centering
        \includegraphics[width=\textwidth]{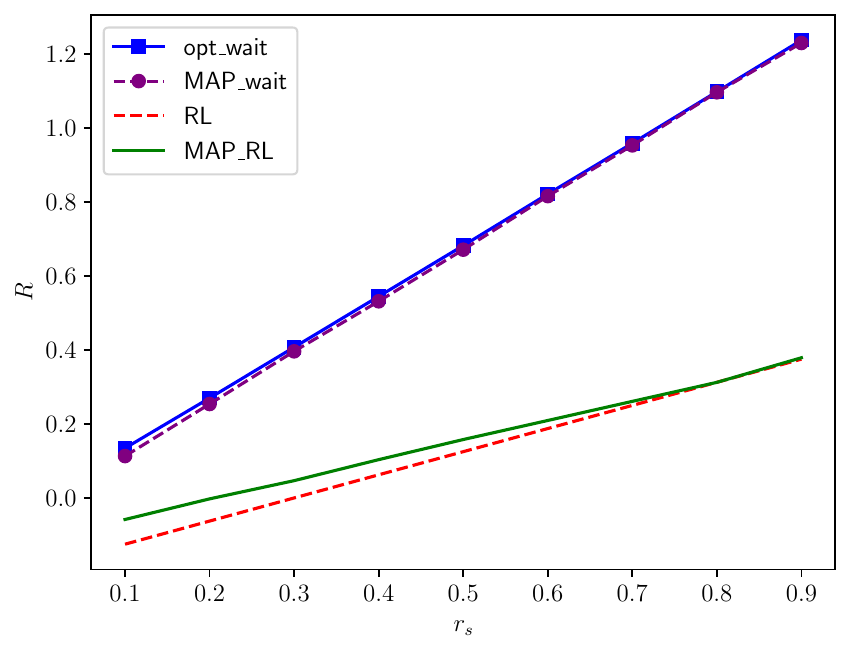}
        \caption{$\alpha=0.5$, $\beta=0.3$.}
        \label{fig:var_rs_high_alpha}
    \end{subfigure}
    \vspace*{0.2cm}
    \caption{Variation of average revenue per unit job $R$ with $r_s$ for $\mu=0.4$, $\lambda=0.3$ and $c_d=3$.}
    \label{fig:var_rs}
\end{figure*}

In the third experiment, we evaluate the behavior of these job submission policies as we vary $r_s$ which is the revenue generated per successful job completion. As shown in Fig.~\ref{fig:var_rs}, again we see that \emph{opt\_wait} policy is superior to the benchmark policies in all cases.

In the first experiment, we note that the RL policy is generally better than the MAP\_RL policy at larger values of $\mu$, for the MM with $\alpha<\beta$ and the opposite trend is observed for the MM with $\beta>\alpha$. This is because, when $\alpha<\beta$, the MM is more likely to be free and higher sampling rate leads to a fresher estimate. Thus, submitting jobs only when $\hat{X}(t)=0$ leads to lower risk.  On the other hand, across all the experiments, we see that MAP\_wait underperforms by a considerable margin when $\alpha<\beta$. The reason for this big disparity in the MAP\_wait policy for the two MMs, is due to the fact that it submits jobs right at the switching point of the MAP estimate in the case of $\alpha\leq \beta$ and $\hat{X}(t)=1$. Hence, the risk associated with getting a penalty is somewhat significant at this point. Therefore, the wait times must be carefully crafted to maximize the return.

\section{Conclusion}
In this paper, we considered a resource allocator (RA) and a Markov machine (MM). The MM oscillates between free and busy depending on its internal dynamics. The RA keeps track of the state of the MM by sampling it, and submits arriving jobs to the MM so as to maximize the revenue. We found the optimal job submission policy for the RA, which has a unit-sized buffer to keep an incoming job until its submission to the MM. We showed that there is a significant performance gain when we carefully select when to submit jobs based on the RA's estimate of the state of the MM and the age of its estimate. The optimal sampling rate allocation problem for monitoring multiple MMs under the optimal job submission policy is left for future work. Further, finding the jointly optimal sampling and job submission policy is another interesting extension to this line of research.

\appendices
\section{Proof of Theorem \ref{thrm:avg_rev}}\label{apen:thrm_avg_rev}
Let $N_i(t)$ be the number of jobs accepted by the RA when $\hat{X}(t)=i$. Once a job is accepted, it  will undergo a semi-Markovian reward process until the job is submitted to the MM. Let us denote this reward process as $M_t(i,u,\tau_{i,u})$ if the job was accepted when $\hat{X}(t)=i$ and $\Delta=u$. Then, this reward process will evolve as follows: Starting from state $(i,u,\tau_{i,u})$,  if the RA does not receive a new update by  $\tau_{i,u}$ time units, the RA submits the job to the MM upon which the job may be discarded or accepted by the MM depending on its current state. However, if the RA receives a new status update before, $\tau_{i,u}$ time units have elapsed, then it transitions to state $(j,0,\tau_{j,0})$ where $j$ is the sampled state of $X(t)$. Then, the RA waits for $\tau_{j,0}$ time units before submitting the job. If a new sample arrives during the waiting period, then state transitions to $(k,0,\tau_{k,0})$ where $k$ is the state of $X(t)$ reported by the new sample. Then, it waits for $\tau_{k,0}$ time units before submitting a job. This process is repeated until the job is submitted to the MM upon which the reward process $M_t(i,u,\tau_{i,u})$ will terminate generating a revenue or penalty.

Let $r(i,u,\tau_{i,u})$ be the reward received at the end of the $M_t(i,u,\tau_{i,u})$. Let $l(i,u,\tau_{i,u})$ be the number of jobs lost to the RA while the reward process was in progress. Now, to find the average revenue we can partition the jobs that arrive at the RA based on the reward processes that are initiated at the RA. For example, when a job arrives at the RA for the first time, the buffer is empty, and hence, the job is accepted by the RA. Suppose when this job is accepted by the RA, $\hat{X}(t)=0$ and $\Delta=u_1$. Then, this will initiate the reward process $M_t^1(0,u_1,\tau_{0,u_1})$ resulting in a reward of $r_1(0,u_1,\tau_{0,u_1})$. During this reward process, the number of jobs lost at the RA is denoted by $l_1(0,u_1,\tau_{0,u_1})$. Hence, at the end of the reward process, the total number jobs received will be equal to $l_1(0,u_1,\tau_{0,u_1})+1$ including the job that initiated the reward process. After  $M_t^1(0,u_1,\tau_{0,u_1})$  has terminated, a new reward process will begin upon the arrival of a job request at the RA. Similarly, all the jobs arriving at the RA can be partitioned based on the reward processes. Let $r_m(i,u_m,\tau_{i,u_m})$ be the reward received at the end of the $m$th reward process that started at the RA when $\hat{X}(t)=i$. In here, $u_m$ is the age of the estimator when this reward process was initiated. Let $l_m(i,u_m,\tau_{i,u_m})$  denote the number of jobs lost at the RA during the same reward process. Then, $R$ can be obtained as follows,
\begin{align}
   R&=\lim_{t\to\infty} \frac{r_sS(t)-c_dD(t)}{N(t)}\\
   &=\lim_{t\to\infty}\frac{\sum_{i=0}^1\sum_{m=1}^{N_i(t)}r_m(i,u_m,\tau_{i,u_m})}{\sum_{i=0}^1\sum_{m=1}^{N_i(t)}1+l_m(i,u_m,\tau_{i,u_m})} \\
    &=\lim_{t\to\infty}\frac{\sum_{i=0}^1 \frac{N_i(t)}{N_0(t)+N_1(t)}\frac{1}{N_i(t)}\sum_{m=1}^{N_i(t)}r_m(i,u_m,\tau_{i,u_m})}{\sum_{i=0}^1\frac{N_i(t)}{N_0(t)+N_1(t)}\frac{1}{N_i(t)}\sum_{m=1}^{N_i(t)}1+l_m(i,u_m,\tau_{i,u_m})}
\end{align}
Since $N_i(t)\to \infty$ as $t\to \infty$, we have that $\frac{N_i(t)}{N_1(t)+N_0(t)}\to p_i$, $\frac{1}{N_i(t)}\sum_{m=1}^{N_i(t)}r_m(i,u_m,\tau_{i,u_m})
\to \e[R(i,U,\tau_{i,U})]$ and $\frac{1}{N_i(t)}\sum_{m=1}^{N_i(t)}l_m(i,u_m,\tau_{i,u_m})\to \e[L(i,U,\tau_{i,U})]$  as $t\to \infty$, where $U$ is the random variable associated with the age of the estimator when a new job is accepted at the RA. Moreover, since after every sampling instance the age of the estimator is reset to zero, the distribution of $U \sim Exp(\lambda+\mu)$.

\section{Proof of Theorem \ref{thrm:opt_wait}}\label{apen:thrm_opt_wait}
Note that $p_i$s are independent of our decision process and simply depend on the dynamics of the underlying CTMC of the query-based sampling procedure (see Remark \ref{rem:pi}). Therefore, to maximize $J(\theta)$, it is sufficient to maximize $\e[V_\theta(i,U,\tau_{i,U})]$ for $i=0,1$. Moreover, if there exists a set of waiting times that jointly maximizes $\e[V_\theta(i,U,\tau_{i,U})|U=u]$ for all $u\geq0$ and $i=0,1$, then those waiting times are optimal for $J(\theta)$.

Let $V_\theta(i,\tau_{i,u},u)=\e[V_\theta(i,U,\tau_{i,U})|U=u]$. Then, by conditioning on the first sampling time which we denote by the random variable $Y$, we can find $V_\theta(i,u,\tau_{i,u})$ as follows,
\begin{align}
    V_\theta&(i,u,\tau_{i,u})\nonumber\\
    =&\e\left[(r_sP_{i0}^{u+\tau_{i,u}}-c_dP_{i1}^{u+\tau_{i,u}}-\lambda\theta \tau_{i,u})\mathds{1}\{Y>\tau_{i,u}\}\right]\nonumber\\
    &+\e\Bigg[\Big(-\lambda\theta Y+ V_\theta(0,0,\tau_{0,0})P_{i0}^{u+Y}\nonumber\\
    &\qquad\quad+V_\theta(1,0,\tau_{1,0})P_{i1}^{u+Y}\Big)\mathds{1}\{Y\leq \tau_{i,u}\}]\\
    =& \Big(r_sP_{i0}^{u+\tau_{i,u}}-c_dP_{i1}^{u+\tau_{i,u}}-\lambda\theta \tau_{i,u}\Big)\e[\mathds{1}\{Y>\tau_{i,u}\}]\nonumber\\
    &+\sum_{j=0}^1\e\Big[P_{ij}^{u+Y}\mathds{1}\{Y\leq \tau_{i,u}\}\Big]V_\theta(j,0,\tau_{j,0})\nonumber\\
    &-\lambda\theta\e[Y\mathds{1}\{Y\leq \tau_{i,u}\}]\\
    =&\Big(r_sP_{i0}^{u+\tau_{i,u}}-c_dP_{i1}^{u+\tau_{i,u}}-\lambda\theta \tau_{i,u}\Big)e^{-\mu\tau_{i,u}}\nonumber\\
    &-\lambda\theta\Big(\frac{1-e^{-\mu\tau_{i,u}}}{\mu}-\tau_{i,u}e^{-\mu\tau_{i,u}}\Big)\nonumber\\
    &+\sum_{j=0}^1\e\Big[P_{ij}^{u+Y}\mathds{1}\{Y\leq \tau_{i,u}\}\Big]V_\theta(j,0,\tau_{j,0})\\
    =&\Big(r_sP_{i0}^{u+\tau_{i,u}}-c_dP_{i1}^{u+\tau_{i,u}}\Big)e^{-\mu\tau_{i,u}}-\frac{\lambda\theta(1-e^{-\mu\tau_{i,u}})}{\mu}\nonumber\\
    &+\sum_{j=0}^1\e\Big[P_{ij}^{u+Y}\mathds{1}\{Y\leq \tau_{i,u}\}\Big]V_\theta(j,0,\tau_{j,0}).\label{eqn:gen_recur}
\end{align}
Moreover, by setting $u=0$ in  \eqref{eqn:gen_recur} we have that,
\begin{align}
    V_\theta(i,0,\tau_{i,0})=&\Big(r_sP_{i0}^{\tau_{i,0}}-c_dP_{i1}^{\tau_{i,0}}\Big)e^{-\mu\tau_{i,0}}-\frac{\lambda\theta(1-e^{-\mu\tau_{i,0}})}{\mu}\nonumber\\
    &+\sum_{j=0}^1\e\Big[P_{ij}^{Y}\mathds{1}\{Y\leq \tau_{i,0}\}\Big]V_\theta(j,0,\tau_{j,0})\label{eqn:Vi_rec}
\end{align}
Note that \eqref{eqn:Vi_rec} depends only on the waiting times $\tau_{0,0}$ and $\tau_{1,0}$. Hence, $V(0,0,\tau_{0,0})$ and $V(1,0,\tau_{1,0})$ can be maximized independently from the rest of our waiting decisions. Further, we have that $V_\theta(0,0,\tau_{0,0})=R(0,0,\tau_{0,0})-\theta L(0,0,\tau_{0,0})\leq R(0,0,\tau_{0,0})\leq r_s$. Therefore, in \eqref{eqn:Vi_rec}, for $i=0$, if we set $\tau_{0,0}=0$, we get that $V_\theta(0,0,0)=r_s$. Hence, $V_\theta(0,0,\tau_{0,0})$ is maximized at $\tau_{0,0}=0$ regardless of our choice for $\tau_{1,0}$. At the same time, by setting $i=1$ in \eqref{eqn:Vi_rec}, we see that maximizing $V_\theta(0,0,\tau_{0,0})$ would also maximize $V_\theta(1,0,\tau_{1,0})$ as well. Hence, the choice of $\tau^*_{0,0}=0$ is jointly optimal for $V_\theta(1,0,\tau_{1,0})$ as well. Now, from \eqref{eqn:Vi_rec}, we can find $\tau_{1,0}$ as follows,
\begin{align}
    V_\theta(1,0,\tau_{1,0})=&\Big(r_sP_{10}^{\tau_{1,0}}-c_dP_{11}^{\tau_{1,0}}\Big)e^{-\mu\tau_{1,0}}-\frac{\lambda\theta(1-e^{-\mu\tau_{1,0}})}{\mu}\nonumber\\
    &+\e\Big[P_{10}^{Y}\mathds{1}\{Y\leq \tau_{1,0}\}\Big]V_\theta(0,0,0)\nonumber\\
    &+\e\Big[P_{11}^{Y}\mathds{1}\{Y\leq \tau_{1,0}\}\Big]V_\theta(1,0,\tau_{1,0}).\label{eqn:v1_tau_usmp}
\end{align}  
Rearranging \eqref{eqn:v1_tau_usmp} gives us,
\begin{align}
V_\theta(1,0,\tau_{1,0})=&\frac{\Big(r_sP_{10}^{\tau_{1,0}}-c_dP_{11}^{\tau_{1,0}}\Big)e^{-\mu\tau_{1,0}}-\frac{\lambda\theta(1-e^{-\mu\tau_{1,0}})}{\mu}}{1-\e\Big[P_{11}^{Y}\mathds{1}\{Y\leq \tau_{1,0}\}\Big]}\nonumber\\
     &+\frac{r_s\e\Big[P_{10}^{Y}\mathds{1}\{Y\leq \tau_{1,0}\}\Big]}{1-\e\Big[P_{11}^{Y}\mathds{1}\{Y\leq \tau_{1,0}\}\Big]}.\label{eqn:v1_max_2}
\end{align}
Note that the above expression for $V_\theta(1,0,\tau_{1,0})$ only depends on $\tau_{1,0}$ and hence it can be maximized independently of other decisions. Let $V_1=\max_{\tau_{1,0}\geq0}V_\theta(1,0,\tau_{1,0})$ and $\tau^*_{1,0}$ be its maximizer. Also for brevity, let $V_0=V(0,0,0)$. From \eqref{eqn:gen_recur}, we have that maximizing $V(0,0,\tau_{0,0})$ and $V(1,0,\tau_{1,0})$ maximizes $V(i,u,\tau_{i,u})$ for all $u>0$ regardless of whether $i=0$ or $i=1$. Hence, $\tau^*_{0,0}$ and $\tau^*_{1,0}$ are jointly optimal for maximizing $V_\theta(i,u,\tau_{i,u})$ for all $u$. This yields that  $\tau^*_{i,u}$ is simply the maximizing argument of \eqref{eqn:gen_recur} when $\tau_{i,0}=\tau^*_{i,0}$. This completes the proof.

\section{Proof of Theorems \ref{cor:opt_times_2} \& \ref{cor:opt_times_3}}\label{apen:cor_opt_times}
We begin the proof using the following observation. Since $V_1=\max_{\tau\geq0}V(1,0,\tau_{1,0})$, we have that $V_1\geq V(1,0,0)=-c_d$. This gives us that $B\geq 0$. Let $f_0^u(\tau)=Ae^{-\mu\tau}+e^{-(\alpha+\beta)u}B_0e^{-(\alpha+\beta+\mu)\tau}$ and $f_1^u(\tau)=Ae^{-\mu\tau}-e^{-(\alpha+\beta)u}B_1e^{-(\alpha+\beta+\mu)\tau}$. Then, $\tau^*_{0,u}=\argmax_{\tau\geq0}f_0^u(\tau)$ and $\tau^*_{1,u}=\argmax_{\tau\geq0}f_1^u(\tau)$. The first derivatives of the two functions are given by,
\begin{align}
    \dv{f_0^u(\tau)}{\tau}=-\mu Ae^{-\mu \tau}-(\alpha+\beta+\mu)e^{-(\alpha+\beta)u}B_0e^{-(\alpha+\beta+\mu)\tau},\label{eqn:df_0}\\
    \dv{f_1^u(\tau)}{\tau}=-\mu Ae^{-\mu \tau}+(\alpha+\beta+\mu)e^{-(\alpha+\beta)u}B_1e^{-(\alpha+\beta+\mu)\tau}.\label{eqn:df_1}
\end{align}

Now, we analyze the behavior of the two functions for the cases $A>0$ and $A\leq0$, when $u>0$.
\begin{enumerate}\itemsep1em
    \item \textbf{Case 1 ($\bm{A>0}$):} Since $B\geq 0$ and $A>0$, $f_0^u(\tau)$ is monotonically decreasing with $\tau$. Therefore, $\tau^*_{0,u}=0$. From \eqref{eqn:df_1}, note that if $(\alpha+\beta+\mu)e^{-(\alpha+\beta)u}B_1\leq\mu A$, then $\dv{f_1^u(\tau)}{\tau}\leq0$ since $e^{-(\alpha+\beta+\mu)\tau}$ decays faster than $e^{-\mu \tau}$. Therefore, $f_1^u(\tau)$ is maximized at $\tau=0$ in this case. If $(\alpha+\beta+\mu)e^{-(\alpha+\beta)u}B_1>\mu A$, then $\dv{f_1^u(\tau)}{\tau}>0$ at first and would eventually become negative and will remain negative till $\tau\to \infty$. Therefore, the function $f_1^u(\tau)$ will be maximized at some $\tau>0$. By setting $\dv{f_1^u(\tau)}{\tau}=0$ yields this value to be $\frac{1}{\alpha+\beta}\ln{\frac{(\alpha+\beta+\mu)B_1}{\mu A}}-u$ which is positive in this scenario.
    
    \item \textbf{Case 2 ($\bm{A\leq0}$):} When $A=0$, then $f_0^u(\tau)$ is monotonically decreasing and $f_0^u(\tau)$ is monotonically increasing. Thus, in this case $\tau^*_{0,u}=0$ and $\tau^*_{1,u}=\infty$. Now, consider the case where $A<0$. Since $B>0$ and $A<0$, we have that $f_1^u(\tau)<0$ and it monotonically increases as $\tau \to \infty$. Therefore, $\tau^*_{1,u}=\infty$. From \eqref{eqn:df_0}, note that, if $(\alpha+\beta+\mu)e^{-(\alpha+\beta)u}B_0\leq-\mu A$, then $\dv{f_0^u(\tau)}{\tau}\geq0$ since $e^{-(\alpha+\beta+\mu)\tau}$ decays faster than $e^{-\mu \tau}$. Therefore, $f_0^u(\tau)$ is maximized at $\tau=\infty$. If $(\alpha+\beta+\mu)e^{-(\alpha+\beta)u}B_0>-\mu A$, then $\dv{f_1^u(\tau)}{\tau}<0$ at first and would eventually become positive and will remain positive till $\tau\to \infty$. This implies that the maximum of $f_0^u(\tau)$ is achieved at $\tau=0$ or at $\tau=\infty$. In particular, if $f^u_0(0)> 0$ then the maximum is achieved at zero, and otherwise at infinity since $f^u_0(\infty)=0$. Now, all these conditions can be concisely represented as follows. If $f^u_0(0)\leq 0$, then regardless whether $(\alpha+\beta+\mu)e^{-(\alpha+\beta)u}B_0\leq-\mu A$ or $(\alpha+\beta+\mu)e^{-(\alpha+\beta)u}B_0>-\mu A$, we will have $\tau^*_{0,u}=\infty$. On the other hand, if $f^u(0)>0$, we have that $(\alpha+\beta+\mu)e^{-(\alpha+\beta)u}B_0>-\mu A$. Thus, in this case $\tau^*_{0,u}=0$.
\end{enumerate}

Next, we will show that our optimal waiting time equations still hold, even at $u=0$. When $u=0$, regardless of $A$, $\tau^*_{0,0}=0$ as we have shown in Appendix \ref{apen:thrm_opt_wait}. Now, to analyze $\tau^*_{1,0}$,  from \eqref{eqn:tau_1} and \eqref{eqn:v1_max_2}, we have that $V_1=\max_{\tau\geq0}V_\theta(1,0,\tau)$. For brevity, let $V_\tau=V_\theta(1,0,\tau)$ and  let  $L(\tau,\gamma)=V_\tau-\gamma \tau$ denote the Lagrangian function of this optimization problem. Now, since the constraint is linear, the KKT conditions give the following necessary conditions for optimality,
\begin{align}
    \pdv{V_\tau}{\tau}-\gamma&=0,\\
    \gamma\tau&=0.
\end{align}
Therefore, we have that if $\tau>0$, then $\gamma=0$, and therefore, $ \pdv{V_\tau}{\tau}=0$ at the optimal point. Now, suppose that the optimal satisfies $\tau>0$. From \eqref{eqn:v1_tau_usmp}, we know that $V_\tau$ satisfies the following,
\begin{align}
    V_\tau=&A_\tau e^{-\mu\tau}-B_\tau e^{-(\alpha+\beta+\mu)\tau}+B_\tau-A_\tau-c_d\label{eqn:v_tau_pre_diff}
\end{align}
where $A_\tau$ and $B_\tau$ are obtained by replacing $V_1$ with $V_\tau$ in the expressions of $A$ and $B_1$, respectively. Then, differentiating \eqref{eqn:v_tau_pre_diff} with respect to $\tau$ yields,
\begin{align}
   \dv{V_\tau}{\tau}=&\dv{A_\tau}{\tau}e^{-\mu\tau}-\mu A_\tau e^{-\mu\tau}-\dv{B_\tau}{\tau}e^{-(\alpha+\beta+\mu)\tau}\nonumber\\
   &+(\alpha+\beta+\mu)B_\tau e^{-(\alpha+\beta+\mu)\tau}+\dv{B_\tau}{\tau}-\dv{A_\tau}{\tau}
\end{align}
Now, since $\dv{V_\tau}{\tau}=0$, we have that $\dv{A_\tau}{\tau}=\dv{B_\tau}{\tau}=0$ since they are linear functions of $V_\tau$. This gives us that, at the optimal point, the following equation should be satisfied,
\begin{align}
    e^{-\mu\tau}(\mu A_\tau-(\alpha+\beta+\mu)B_\tau e^{-(\alpha+\beta)\tau})=0.\label{eqn:opt_cond}
\end{align}
Therefore, from \eqref{eqn:opt_cond} and the KKT conditions, we have that $\tau^*_{1,0}$ must be either $0$, $\infty$ or $\Gamma$. In here, $\Gamma$ is the value of $\tau$ at which $\mu A_\tau=(\alpha+\beta+\mu)B_\tau e^{-(\alpha+\beta)\tau}$ is satisfied. When $\tau=0$, we note that $V_\tau=-c_d$. Therefore, if $A\leq 0$, note that \eqref{eqn:opt_cond} can only be satisfied at $\tau=\infty$ since $B>0$. Further, since $A\leq0$, we have that $0<\frac{\lambda\theta}{\mu}\leq\frac{\alpha}{\alpha+\beta}(V_1+c_d)$. Therefore, we have that $V_1>-c_d$. This gives us that $\tau^*_{0,1}=\infty$, in this particular case.

Now, consider the case where $A>0$ and suppose $\Gamma>0$. We see that, from \eqref{eqn:v_tau_pre_diff}, at $\tau=\infty$ we have that $V_\infty=r_s-\frac{\lambda\theta(\alpha+\beta+\mu)}{\mu\beta}$ and $V_\Gamma=\frac{(\alpha+\beta)A_\Gamma}{\beta}e^{-\mu\Gamma}+r_s-\frac{\lambda\theta(\alpha+\beta+\mu)}{\mu\beta}>V_\infty.$ Alternatively, $V_\Gamma$ can be written as $V_\Gamma=\frac{(\alpha+\beta)A_\Gamma e^{-\mu\Gamma}}{(\alpha+\beta+\mu)}+B_\Gamma-A_\Gamma-c_d$. Now, for $\Gamma>0$, we have,
\begin{align}
    \frac{e^{(\alpha+\beta)\Gamma}-1}{\alpha+\beta}>\Gamma> \frac{1-e^{-\mu\Gamma}}{\mu},\label{eqn:ineq_gamma}
\end{align}
where the last inequality is obtained using the fact that $x> 1-e^{-x}$ for $x>0$. Now, substituting for $\Gamma$ in \eqref{eqn:ineq_gamma}, we get,
\begin{align}
    \frac{(\alpha+\beta+\mu)B_\Gamma}{\mu A_\Gamma}>&1+\frac{\alpha+\beta}{\mu}(1-e^{-\mu\Gamma})\\
    B_\Gamma>&A_\Gamma-\frac{\alpha+\beta}{\alpha+\beta+\mu}A_\Gamma e^{-\mu\Gamma}.
\end{align}
This gives us that $V_\Gamma>-c_d$. Hence, $\tau^*_{1,0}=\Gamma$. If $\Gamma\leq0$, then we only have to compare $V_\tau$ at $\tau=0$ and $\tau=\infty$. Since $\Gamma=0$, we have that at the optimal $\tau$, $\mu A_\tau\geq (\alpha+\beta+\mu)B_\tau$. This implies that at the optimal $\tau$, $A_\tau>B_\tau$. Therefore, if $\tau=\infty$ is optimal, then we will have $V_\infty<-c_d$ which is a contradiction. Therefore, the optimal must be achieved at $\tau=0$. Hence, $\tau^*_{1,0}=0$ in this case.
This concludes the proof.

\section{Proof of Lemma \ref{lem:avg_rev_pos}}\label{apen:lem_avg_rev_pos}
When $A>0$, we have that $\tau^*_{0,u}=0$. Substituting this in \eqref{eqn:gen_recur} yields,
\begin{align}
    V_\theta(0,u,0)&=r_sP_{00}^u-c_dP_{01}^u\\
    &=\frac{\beta r_s-\alpha c_d}{\alpha+\beta}+\frac{\alpha (r_s+c_d)}{\alpha+\beta}e^{-(\alpha+\beta)u}\label{eqn:v0_t0}
\end{align}
Then, taking expectation with respect to $U$ gives the first part of the result. To prove the next part, first consider the case in which $u>\Gamma$. In this case, $\tau^*_{1,u}=0$. Substituting this in \eqref{eqn:gen_recur}, gives us that,
\begin{align}
     V_\theta(1,u,0)=&r_sP_{10}^u-c_dP_{11}^u\\
    =&\frac{\beta r_s-\alpha c_d}{\alpha+\beta}-\frac{\beta (r_s+c_d)}{\alpha+\beta}e^{-(\alpha+\beta)u}
\end{align}
When $u\leq \Gamma$, we have that $\tau^*_{1,u}=\Gamma-u$. Moreover, \eqref{eqn:gen_recur} can be rearranged as follows,
\begin{align}
    V_\theta(1,u,\tau)=&f_1^u(\tau)+\frac{r_s\beta-\alpha c_d}{\alpha+\beta}-A\nonumber\\
    &-\frac{\beta \mu(r_s-V_1)}{(\alpha+\beta)(\alpha+\beta+\mu)}e^{-(\alpha+\beta)u},\label{eqn:v1_tau_gen}
\end{align}
where $f_1^u(\tau)$ is as defined in Appendix \ref{apen:cor_opt_times}. Therefore, we have,
\begin{align}
    V_\theta(1,u,\Gamma-u)=&f_1^u(\Gamma-u)+\frac{r_s\beta-\alpha c_d}{\alpha+\beta}-A\nonumber\\
    &-\frac{\beta \mu(r_s-V_1)}{(\alpha+\beta)(\alpha+\beta+\mu)}e^{-(\alpha+\beta)u}\\
    =&\frac{(\alpha+\beta)A}{\alpha+\beta+\mu}e^{\mu(u-\Gamma)}+\frac{r_s\beta-\alpha c_d}{\alpha+\beta}-A \nonumber\\
    &-\frac{\beta \mu(r_s-V_1)}{(\alpha+\beta)(\alpha+\beta+\mu)}e^{-(\alpha+\beta)u}.
\end{align}
Now, the expectations for the two cases can be obtained as follows,
\begin{align}
    \e[&V_\theta(1,U,\tau^*_{1,U})\mathds{1}\{U>\Gamma\}]\nonumber\\
    =&(\frac{r_s\beta-\alpha c_d}{\alpha+\beta}-A)(1-e^{-(\lambda+\mu)\Gamma})\nonumber\\
    &+\frac{(\alpha+\beta)(\lambda+\mu)A}{\lambda(\alpha+\beta+\mu)}(1-e^{-\lambda\Gamma})e^{-\mu\Gamma}\nonumber\\
    &-\frac{\beta \mu(\lambda+\mu)(r_s-V_1)}{(\alpha+\beta)(\alpha+\beta+\mu)(\alpha+\beta+\mu+\lambda)}(1-e^{-(\alpha+\beta+\mu+\lambda)\Gamma})\\
    \e[&V_\theta(1,U,\tau^*_{1,U})\mathds{1}\{U\leq\Gamma\}]\nonumber\\
    =&\frac{r_s\beta-\alpha c_d}{\alpha+\beta}e^{-(\lambda+\mu)\Gamma}\nonumber\\
    &-\frac{\beta(\lambda+\mu) (r_s+c_d)}{(\alpha+\beta)(\alpha+\beta+\mu+\lambda)}e^{-(\alpha+\beta+\mu+\lambda)\Gamma}.
\end{align}
Adding the above two equations  and substituting $e^{-(\alpha+\beta)\Gamma}=\frac{\mu A}{B_1(\alpha+\beta+\mu)}$ gives the second part of the result.
\section{Proof of Lemma \ref{lem:avg_rev_neg}}\label{apen:lem_avg_rev_neg}
To prove the first part of the result, we first consider the case where $u<\kappa$. In this case, $\tau^*_{0,u}=0$. Therefore, using \eqref{eqn:v0_t0}  and taking expectation with respect to $U$, gives us,
\begin{align}
   \e[&V_\theta(0,U,\tau^*_{0,U})\mathds{1}\{U\leq \kappa\}]\nonumber\\
   =&\frac{r_s\beta-\alpha c_d}{\alpha+\beta}(1-e^{-(\lambda+\mu)\kappa})\nonumber\\
   &+\frac{\alpha(\lambda+\mu)(r_s+c_d)}{(\alpha+\beta)(\alpha+\beta+\mu+\lambda)}(1-e^{-(\alpha+\beta+\mu+\lambda)\kappa})
\end{align}
Now, we consider the case where $u>\kappa$. From  \eqref{eqn:gen_recur}, we have that,
\begin{align}
    V_\theta(0,u,\tau)&=f_0^u(\tau)+\frac{r_s\beta-\alpha c_d}{\alpha+\beta}-A\nonumber\\
    &+\frac{\alpha \mu(r_s-V_1)}{(\alpha+\beta)(\alpha+\beta+\mu)}e^{-(\alpha+\beta)u},
\end{align}
where $f_0^u(\tau)$ is as defined in Appendix \ref{apen:cor_opt_times}. For $u>\kappa$, we have that $\tau^*_{0,u}=\infty$. Note that $f_0^u(\infty)=0$. Therefore, we have 
\begin{align}
     \e[&V_\theta(0,U,\tau^*_{0,U})\mathds{1}\{U>\kappa\}]\nonumber\\
     =&(\frac{r_s\beta-\alpha c_d}{\alpha+\beta}-A)e^{-(\lambda+\mu)\kappa}\nonumber\\
    &+\frac{\alpha \mu(\lambda+\mu)(r_s-V_1)}{(\alpha+\beta)(\alpha+\beta+\mu)(\alpha+\beta+\mu+\lambda)}e^{-(\alpha+\beta+\mu+\lambda)\kappa}.
\end{align}
Next, we will prove the second part of the result. In here, we have that $\tau^*_{1,u}=\infty$. Therefore, using \eqref{eqn:v1_tau_gen} and the fact that $f_1^u(\infty)=0$, we get,
\begin{align}
    V_\theta(1,u,\infty)=&\frac{r_s\beta-\alpha c_d}{\alpha+\beta}-A\nonumber\\
    &-\frac{\beta \mu(r_s-V_1)}{(\alpha+\beta)(\alpha+\beta+\mu)}e^{-(\alpha+\beta)u}.
\end{align}
Then, taking expectation with respect $U$, yields the second part of the result.

\section{Proof of Lemma \ref{lem:pi}}\label{apen:lem_pi}
Let $\tilde{P}(t)$ denote the transition probabilities of $Z(t)$. Then, $\tilde{P}(t)=e^{tQ}$. Now, we note that,
\begin{align}
    Q^n=\begin{bmatrix}
        \Lambda^n&\Lambda^{n-1}\Psi\\
        \bm{0}_1&\bm{0}_2
    \end{bmatrix}.
\end{align}
This give us, 
\begin{align}
    \tilde{P}(t)&=I+Qt+\frac{Q^2t^2}{2!}+ \dots\\
    &=I+\begin{bmatrix}
        \Lambda &\Psi\\
        \bm{0}_1&\bm{0}_2
    \end{bmatrix}t+
    \begin{bmatrix}
        \Lambda^2 &\Lambda\Psi\\
        \bm{0}_1&\bm{0}_2
    \end{bmatrix}\frac{t^2}{2!}+\dots\\
    &=\begin{bmatrix}
        e^{\Lambda t}&(e^{\Lambda t}-I)\Lambda^{-1}\Psi\\
        \bm{0}_1&\bm{0}_2
    \end{bmatrix}.
\end{align}
Now, note that at $t\to \infty$, probability that we are in one of the transient states goes to zero. Therefore, we have that $\lim_{t\to\infty}e^{\Lambda t}=\bm{0}_3$ where $\bm{0}_3$ is a $4\times 4$ matrix of zeros. This yields that,
\begin{align}
   \tilde{P}(\infty)=\begin{bmatrix}
       \bm{0}_3&-\Lambda^{-1}\Psi\\
        \bm{0}_1&\bm{0}_2
    \end{bmatrix}.
\end{align}
Then, multiplying with the initial probability vector will give us the desired result.

\newpage
\bibliographystyle{unsrt}
\bibliography{refs}

\end{document}